\documentclass[aps,prc,superscriptaddress,nofootinbib]{revtex4-2}

\usepackage{amsmath,amssymb}
\usepackage{graphicx}
\usepackage{bm}
\usepackage{array}
\newcolumntype{P}[1]{>{\centering\arraybackslash}p{#1}}
\begin{document}

\title{Determination of the Charge of Electron Using Faraday's Law of Electrolysis}

\author{Vladimir Ya. Kezerashvili}
\affiliation{New York City College of Technology, The City University of New York, Brooklyn, NY, USA}

\author{Roman Ya. Kezerashvili}
\affiliation{New York City College of Technology, The City University of New York, Brooklyn, NY, USA}
\affiliation{The Graduate School and University Center, The City University of New York, New York, NY, USA}
\affiliation{Long Island University, Brooklyn, NY, USA}

\date{September 23, 2026}

\begin{abstract}

We propose a simple experiment for measuring the charge of the electron based on Faraday's law of electrolysis, assuming the known value of Avogadro's number. The experimental setup consists of electrolytic cells connected in series, so that the same electric current passes through each electrolyte. By measuring the mass deposited at the electrodes and relating it to the total electric charge transferred through the electrolyte using Faraday's law, the charge of the electron can be determined. The experiment requires only standard laboratory equipment, and its performance and analysis are within the skills and knowledge of most students taking algebra-based general physics courses. It is therefore particularly suitable for instructional laboratories in high schools and two- and four-year colleges. We demonstrate that students can obtain a reasonable value for the charge of the electron from simple measurements, providing a direct experimental connection between Faraday's law, Avogadro's number, and the electron charge.

\end{abstract}
\maketitle

In 1897 J.J. Thomson at Cambridge University performed one of the landmark
experiments in physics, the measurement of charge-to-mass ratio, e/m, for
the electron \cite{1}. In early 1897, a rough estimate of the electric
charge per carrier was made by Townsend \cite{2}, who studied the
electrification of gases which are given off when a liquid is decomposed by
an electric current. He found that these gases possess the remarkable
property of forming a dense clouds of water droplets when they meet
moisture. In December 1898, Thomson \cite{3} made another estimate of the
elementary charge by forming charged water droplets by $X-$rays ionizing air
saturated with water. The rate of fall of the droplets under gravity and
application of Stokes' law helped to determine the charge. However, the
distinction of being the first precise measurement of the electron charge
goes to Robert Millikan through his oil-drop experiment in 1909 \cite{4,5,6}%
. The measurement of the electron's charge achieved by Millikan with
Thomson's results for the charge-to-mass ratio for the electron, a value for
the electron mass was obtained. This exciting breakthrough, which became
known as the oil-drop experiment, was a major contribution to physics and
Millikan received worldwide recognition for the experiment, including the
Nobel Prize in Physics in 1923. Most science educators would consider
Thomson's cathode rays and Millikan's oil drop experiments to be two of the
most important contributions to our understanding of modern physics.

The number of molecules or atoms in a gram-mole of any substance has been
found to be $N_{0}=6.02\times 10^{23}$ mol$^{-1}$, known as Avogadro's
constant or Avogadro number. The first undeniably reliable measurements of
Avogadro's number came right at the turn of the twentieth century, with
Millikan's measurement of the charge of the electron. In 1913 Millikan using
the value of the elementary electrical charge reported the Avogadro constant 
\cite{7}. What is most impressive is that the values of the Avogadro
constant measured from completely unrelated techniques agree very well with
each other confirming the hypothesis postulated nearly two centuries ago.
The $X-$ray diffraction method provided precision measurement for the
Avogadro number. The densities and isotopic-abundance ratios measurements of
nearly perfect Si single crystals combined with optical interferometry of
the crystal repeat distance, yield a new value for the Avogadro constant: $%
N_{0}=6.0220943\times 10^{23}$ per mole \cite{8} and currently is the best
experimental value. In 2019 \cite{9} the Avogadro constant was one of the
seven constants chosen in such a way that any unit of the SI can be written
either through a defining constant itself or through products or quotients
of defining constants. The redefinition of the mole in 2019 gives for the
Avogadro constant exact value: $N_{0}=6.02214076\times 10^{23}$ mol$^{-1}$
in the International System of Units, the SI. Below we suggest determining
the charge of election through Faraday's Law of electrolysis using the
given exact value of the Avogadro number.

When the electric current passes through the electrolyte, a chemical
decomposition takes place accompanied by the deposition of negative ions at
the positive electrode (anode) and  positive ions at the negative electrode
(cathode). This process is called electrolysis. Thus, in electrolysis the
negative electrode attracts cations, which combine with the electrons
supplied by the battery (power supply). Similarly, the positive electrode
attracts anions, which replenish the electrons removed by the battery.
Michael Faraday discovered and formulated two laws governing electrolysis in
the 1830's.
\begin{figure}[b]
\centering
\includegraphics[width=10.0cm]{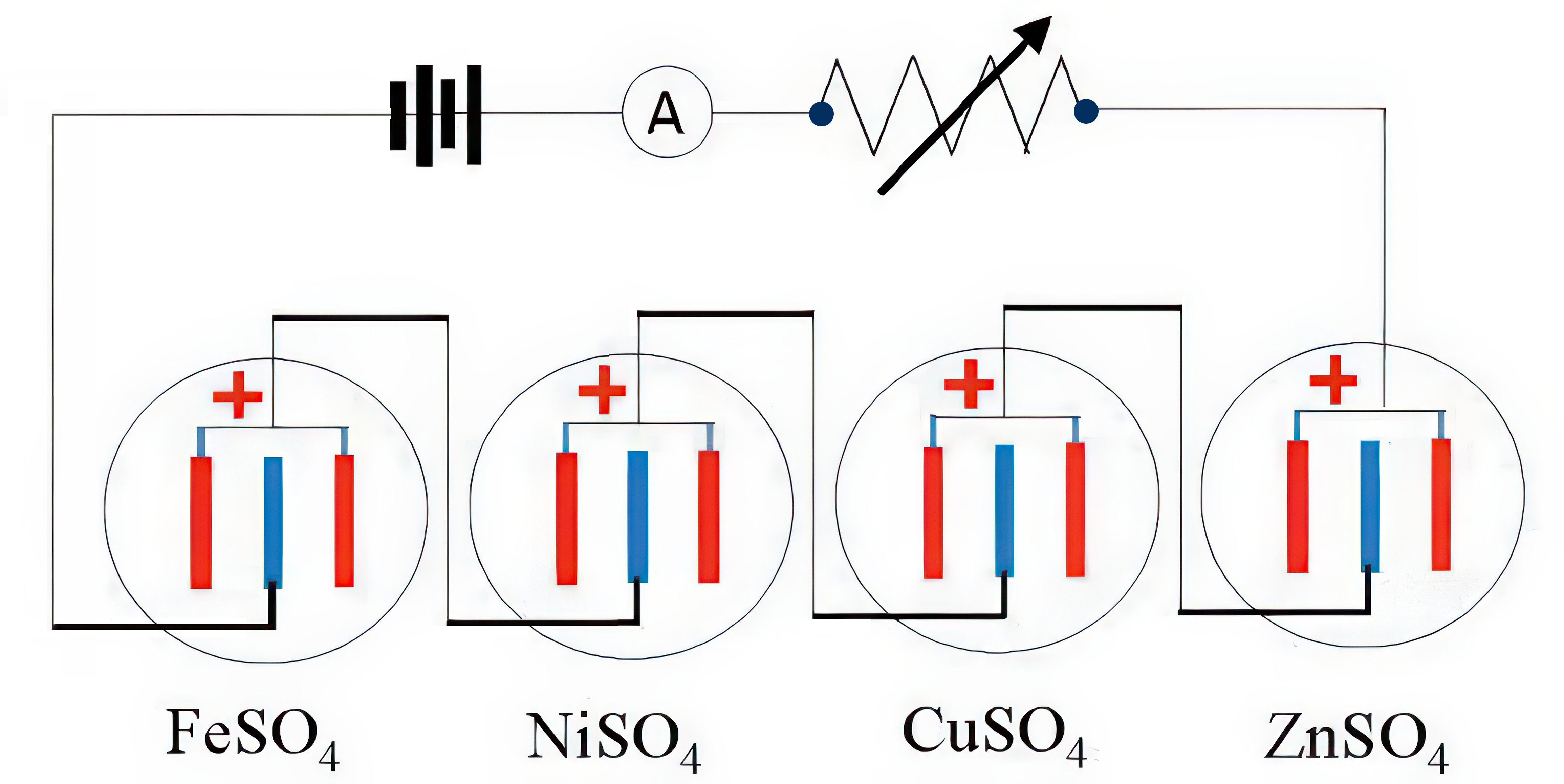}
\caption{(Color online) Schematic circuit diagram for a measurement of the
charge of electron.}
\label{F1}
\end{figure}

For the positive ion to be deposited at the negative electrode, it must
receive the number of electrons equal to its valence. Thus, the charge
necessary to deposit one atom of a substance is $el$, where $e$ is the
charge of a single electron and $l$ is a valence of the element. The valence is considered here as the number of electrons an atom of an element can gain or lose when forming molecules with atoms of other elements. For example, copper in CuSO$_{4}$ has valence 2, which means that Cu lost 2 electrons gained by SO$_{4}$ to form a copper sulfate molecule. The same atom can exhibit different valence depending on the molecules formed. Because
there is Avogadro's number $N_{0}$ of atoms in a mole, the amount of charge
needed to deposit one mole of atoms is $N_{0}el$. But one mole is an amount
of the substance, which mass in grams is numerically equal to its atomic
mass $A$. Therefore, the charge of $N_{0}el$ coulombs will deposit $A$ grams
of the substance, so 1 gram of the substance will plate out $\frac{N_{0}el}{A%
}$ coulombs. Because $q=It$ is the charge in coulombs passed through the
electrolyte, based on Faraday's Law of electrolysis the number of grams of
substance deposited at the electrode is 
\begin{equation}
m=\frac{1}{N_{0}e}\frac{A}{l}It.  \label{Eq1}
\end{equation}%
In Eq. (\ref{Eq1}) $m$ is the mass in grams, $I$ is the current in amperes,
and $t$ is the time in seconds during which the current flowed. Equation (%
\ref{Eq1}) states that the deposited mass $m$ is directly proportional to
the product of the chemical equivalent of the substance, $\frac{A}{l}$, the
electric current, $I$, and time, $t$. The proportionality constant $\frac{1}{%
N_{0}e}$ is a universal constant, being composed only of the fundamental
constants: $N_{0}$ and $e$. Therefore, by measuring the dependence of the
deposit mass as a function of the chemical equivalent of the substance, $%
\frac{A}{l}$, for the same constant electric current and during the same
time we can determine the charge of electron assuming that the Avogadro's
constant $N_{0}$ is known. 

In this experiment, students are using the Faraday's laws of electrolysis to
determine the charge of electron using the fundamental Avogadro constant. To
determine the charge of electron, we suggest using the circuit shown in Fig. %
\ref{F1}. Although one electrolyte would be enough, one can consider four or more jars filled with different electrolytes. When jar cells are connected in series, as shown in Fig. 1, the same current for the same time is flowing in all cells. The use of more than one electrolyte allows us to determine the electron charge as an average for different electrolytes in single experiment and use the spread in the observed values to derive the characteristics of the underlying random error.  The salt solutions are electrolyzed in all jars, and as a result the masses of the elements deposited at the electrodes can be measured and the charge of electron be calculated using Eq. (\ref{Eq1}).
\begin{figure}[t]
\centering
\includegraphics[width=14.0cm]{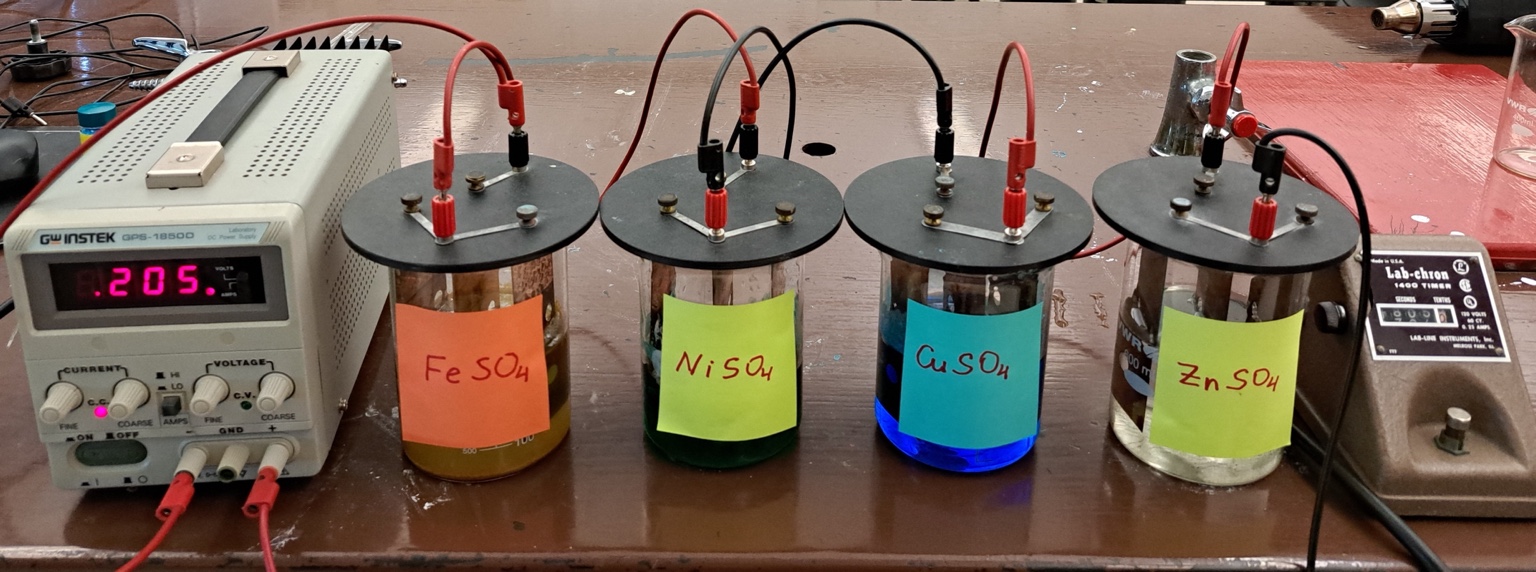}
\caption{(Color online) Experimental setup for measurements of the charge of
electron using the electrolysis.}
\label{F2}
\end{figure}

\emph{Experiment:} The experimental procedure and instructions for analysis
are the following. Designate iron, nickel, copper, and zinc sulfate
electrolytes for each jar. Measure the mass of each cathode plate using an
analytical balance. These plates will be weighed again later to find the
masses deposited on the plates. 
Fill corresponding jars with 20\% by weight 
chemical compounds prepared using the distilled water (1 weight part of a salt and 4 weight parts of distilled water): iron sulfate (FeSO$_{4}$), nickel sulfate (NiSO$_{4}$), copper sulfate
(CuSO$_{4}$) and zinc sulfate (ZnSO$_{4}$) solutions. 
The jars should be filled so that the electrolytic
solution covers about 3/4 of the electrodes.

1. Turn on the power supply, adjust the current through the circuit to 0.200
A, and start the timer. Record the actual value of the current. Allow the
current to flow for about 20-30 minutes. It is important to keep the current
constant for the specified period. Use the power supply in constant current
operation mode or the rheostat to keep the current constant.

2. Carefully remove the previously weighed cathode plates from each jar and
dry them being careful not to remove masses that have been deposited on the
plates. It can take up to 20 minutes to completely dry the plates. A fan can be used to speed up the drying process. The plates are dry when the wet areas are darker in appearance compared to the dry ones once evaporate.  
Weigh and record the masses of the plates.

3. Compute the deposited masses of iron, nickel, copper, and zinc using your
experimental data.

4. Use equation (\ref{Eq1}) to compute the charge of electron $e$ for each
case of the deposited masses $m$. Use the value of the chemical equivalent
of the substance, Avogadro constant, current and the time during which the
current flowed to calculate the electron charge.

5. Compute the average value of the electron charge and compare this value
with standard value of the electric charge by computing the percent error.

The
actual setting for the experiment is shown in Fig. \ref{F2}. We used four jar cells with Cu plates and performed experimental
measurements during $t=1800$ s with the established current $I=0.205$ A. The results of
measurements and calculations are summarized in Table I. 
\begin{table}[t]
\caption{Data for determination of the deposit mass of a metal due to the
electrolysis. The charge of electron is calculated using Eq. (\protect\ref%
{Eq1}). }
\label{tab11}
\begin{center}
\resizebox{\textwidth}{!}{%
\begin{tabular}{cccccccc}
\hline\hline
Substance & Chemical equivalent, & Time, & Current, & Mass before, & Mass
after, & Deposit mass, & Charge of electron, \\ 
& $A/l$, g mol$^{-1}$ & $t$, s & $I$,\ A & $m_{i}$,\ g & $m_{f}$,\ g & $m$, g
& $e$, C \\ \hline
FeSO$_{4}$ & 27.9225 & 1800 & 0.205 & 29.1533 & 29.2560 & 0.1027 & (1.67$\pm 
$0.032)$\times $10$^{-19}$ \\ 
NiSO$_{4}$ & 29.3467 & 1800 & 0.205 & 29.7785 & 29.8884 & 0.1099 & (1.64$\pm 
$0.031)$\times $10$^{-19}$ \\ 
CuSO$_{4}$ & 31.773 & 1800 & 0.205 & 30.5209 & 30.6540 & 0.1331 & (1.46$\pm $%
0.027)$\times $10$^{-19}$ \\ 
ZnSO$_{4}$ & 32.69 & 1800 & 0.205 & 30.1916 & 30.3136 & 0.1220 & (1.64$\pm $%
0.030)$\times $10$^{-19}$ \\ \hline
\multicolumn{8}{c}{Average: $e=1.60\times 10^{-19}$ C; $\ \sigma =8.1\times $%
10$^{-21}$ C. $e=(1.60\pm 0.081)\times 10^{-19}$ C} \\ \hline\hline
\end{tabular}%
}%
\end{center}
\end{table}

The deposit masses of iron, nickel, copper, and zinc were determined as a
difference in masses of the corresponding plates before and after the
electrolysis. Fisher Scientific A-160 analytical balance was used for mass
measurement \cite{10}. Each mass was determined as an average of three
reading at random moments of the stable balance. Although, the mass reading
was done with 0.1 mg precision, the error in mass determination $\Delta{m}$
was 1.0 mg due to mass reading changes of 1.0 mg during prolong observation
of the readings in the balance stable state. GW instek GPS-1850D power
supply \cite{11} was used to establish 0.205 A sustained current in constant
current operational mode. The current readings were continuously monitored
during the experiment duration and no changes of the reading were observed.
The error in the current value was calculated as $\Delta I=I\times 0.002+3$%
mA due to (load regulation) $\leq 0.2\%+3$ mA \cite{11} for constant current
operation mode. Both errors in mass and current are contribution to the
charge error $\Delta e=e \sqrt{(\Delta m/m)^{2} +(\Delta I/I)^{2})}$.  Notice that the electrolysis of water itself is negligible because almost the entire current in the electrolytes is carried by ions of the metal sulfates dissolved into the distilled water, not by the hydrogen and oxygen ions of dissolved water molecules. This is strongly supported by the well-known fact that the conductivity of distilled water is 3 to 6 orders of magnitude less than the conductivity of drinking or salted water. Also, no hydrogen and oxygen gas bubbles associated with water electrolysis were observed at the electrodes in the experiment.

The calculated values of
the electron charge $e$ with the corresponding error are listed in Table 1.
The experimental value of the charge for CuSO$_{4}$ is outside of $3\Delta e$
margin of the standard electron charge value. For three other substances, the
charge value is within the margin of one or two charge errors to the
standard electron charge value. The observed discrepancy for CuSO$_{4}$
signals of the systematic error. Later measurement of the plate with the Cu
mass deposit put back into CuSO$_{4}$ solution shown a noticeable mass
increase without the running current. We did not correct the observed Cu
mass deposit in Table 1 for possible additional mass, because the claim that
the addition mass deposit rate is the same with running current as without it is
unfounded. Finally, we obtained the average value of the electron charge $%
e=(1.60\pm 0.081)\times 10^{-19}$ C, where the $\sigma =8.1\times 10^{-21}$
C is the standard deviation from the average. This is within 5\% margin of
the standard value of the charge of electron. The comparison of the average
value for the electron charge with the standard value $1.602176634\times
10^{-19}$ C leads to the less than 1\% error.  
Therefore, the percentage error for
the charge of electron obtained from suggested simple experimental set that
can be performed within the standard lab classroom period is small enough.
In the classroom, this experiment can be successfully performed with four or
more electrolytes. For this experiment one can use the Wabash Instrument
Corporation (WINSCO) \cite{12} jars and electrode plates are available from
Arbor Scientific (P6-2605) \cite{13}.

\emph{Concluding remarks.} In today's market, some companies offer
educational setups based on the Millikan Oil Drop experiment for measurement of
the charge of electron. For example, the PASCO Millikan Oil Drop Apparatus 
\cite{14} is designed to conduct the Millikan Oil Drop Experiment where the
electric charge carried by a particle may be calculated by measuring the
force experienced by the particle in an electric field of known strength.
The behavior of small mass charged droplets of oil is observed in a
gravitational and an electric field. Measuring the velocity of fall of the
drop in air enables, with the use of Stokes' Law, the calculation of the
mass of the drop. Although this experiment will allow the total charge on a
drop to be measured, it is only through an analysis of the data obtained and
a certain degree of experimental skill that the charge of a single electron
can be determined. The performance and analysis of this experiment are beyond
the scope of skills and knowledge of most students taking Algebra-based
general physics classes. We are suggesting a simple experiment for the
measurement of the charge of an electron based on Faraday's Law of
electrolysis assuming the knowledge of the Avogadro number. Our experience
justifies \cite{Kezerashvili2003Mechanics,Kezerashvili2003Electricity} that for a 1h 40 minutes laboratory class with 24 students working
within 8 groups of 3 students in the group the experimental measurements
take about $45-55$ minutes. The rest of the class time is devoted to an
instructor's introduction to the experiment and calculations and data
analysis.


\begin{thebibliography}{99}
\bibitem{1} J. J. Thomson, Cathode rays. Philosophical Magazine, 44,
293--316 (1897).

\bibitem{2} J. S. Townsend, On electricity in gases and the formation of
clouds in charged gases. Proceedings of the Cambridge Philosophical Society,
9, 244 (1897).

\bibitem{3} J. J. Thomson, On the charge of electricity carried by the ions
produced by R\"{o}ntgen Rays. Philosophical Magazine, 46, 528 (1898).

\bibitem{4} R. A. Millikan, A new modification of the cloud method of
measuring the elementary electrical charge, and the most probable value of
that charge, Proceedings of the American Physical Society. Minutes of the
Forty-Seventh Meeting XXIX (6), 560-561 (1909).

\bibitem{5} R. A. Millikan, A new modification of the cloud method of
determining the elementary electrical charge and the most probable value of
that charge. Philosophical Magazine 19, 209--228 (1910).

\bibitem{6} R. A. Millikan, The isolation of an ion, a precision measurement
of its charge, and the correction of Stokes's law. Phys. Rev. 32, 349-397
(1911).

\bibitem{7} R. A. Millikan, On the elementary electrical charge and the
Avogadro constant. Phys. Rev. 2, 109--143 (1913).

\bibitem{8} R. D. Deslattes, et al., Determination of the Avogadro constant.
Phys. Rev. Lett. 33, 463 (1974).

\bibitem{9} D. B. Newell and E. Tiesinga The International System of Units
(SI). NIST Special Publication 330, National Institute of Standards and
Technology (2019). doi:10.6028/nist.sp.330-2019 S2CID 242934226

\bibitem{10} %
\url{https://www.manualslib.com/manual/1015482/Fisher-Scientific-A-160.html}

\bibitem{11} \url{https://www.gwinstek.com/en-US/products/detail/GPS-Series}

\bibitem{12} \url{https://www.winsco.com/decli/wp-content/uploads/inst_gs432a.pdf}

\bibitem{13} \url{https://www.arborsci.com/products/voltaic-cell-with-electrodes}

\bibitem{14} PASCO Millikan Oil Drop Apparatus, AP-8210A
\url{https://www.pasco.com/products/lab-apparatus/fundamental-constants/ap-8210}

\bibitem{Kezerashvili2003Electricity}
R.~Ya.~Kezerashvili,
\textit{College Physics Laboratory Experiments: Electricity, Magnetism, Optics}
(Gurami Publishing, New York, 2003),
ISBN 978-0-9743929-9-8.

\bibitem{Kezerashvili2003Mechanics}
R.~Ya.~Kezerashvili,
\textit{Laboratory Experiments in College Physics: Mechanics and Heat}
(Gurami Publishing, New York, 2011),
ISBN 978-0-9743929-3-6.
\end{thebibliography}
\end{document}